\documentclass[sn-mathphys]{sn-jnl}

\jyear{2021}%

\theoremstyle{thmstyleone}%
\theoremstyle{thmstyletwo}%

\theoremstyle{thmstylethree}%

\usepackage[utf8]{inputenc}
\usepackage{graphicx}
\usepackage{caption}
\usepackage{epsfig,epsf}
\usepackage{amsmath}
\usepackage{amsthm}
\usepackage{amsfonts}
\usepackage{amssymb}
\usepackage{dsfont}
\usepackage{multirow}
\usepackage{appendix}
\usepackage{slashed}
\usepackage[active]{srcltx}
\usepackage{psfrag}
\usepackage{subfigure}
\usepackage{capt-of}
\usepackage[utf8]{inputenc}
\usepackage{braket}
\usepackage{physics}
\begin{document}

\title[Article Title]{Possible Molecular Explanation for the Resonance $Y(4500)$}


\author[1]{\fnm{E.} \sur{G\"ung\"or}}\email{elif-gungor3011@hotmail.com}
\equalcont{These authors contributed equally to this work.}

\author[1]{\fnm{H.} \sur{Sundu}}\email{hayriye.sundu@kocaeli.edu.tr}
\equalcont{These authors contributed equally to this work.}

\author*[1]{\fnm{J.} \sur{Y. S\"ung\"u}}\email{jyilmazkaya@kocaeli.edu.tr}
\equalcont{These authors contributed equally to this work.}

\author[1]{\fnm{E.} \sur{V.Veliev}}\email{elsen@kocaeli.edu.tr}
\equalcont{These authors contributed equally to this work.}

\affil*[1]{\orgdiv{Physics}, \orgname{Kocaeli University}, \orgaddress{\street{Umuttepe Yerleşkesi}, \city{Izmit}, \postcode{41001}, \country{Turkey}}}


\abstract{The BESIII collaboration has discovered a new state with hidden charm-strange. Its mass is intriguingly close to the $ D_s\bar{D}_{s1} $ threshold and does not have the properties of the charmonium states. Working with the QCD sum rules (QCDSR) approach, we test if the charmonium-like structure $ Y (4500) $, detected in the invariant mass spectrum $ K^+ K^-J/\psi$ may be interpreted as an exotic $ D_s\bar{D}_{s1} $ molecular structure with $ J^{PC}=1^{--} $. Considering the contributions of QCD condensates up to operator dimension ten, we estimate the mass and decay constant of $Y(4500)$ resonance. We get $ m_Y = (4488.35 \pm 11.54) $ MeV in excellent agreement with the meson mass reported by BESIII and $ f_Y=(4.04 \pm 0.36)\times 10^{-3} $$ \mathrm{GeV^4} $.  Our findings indicate that a pseudoscalar-axialvector molecule current can well describe this state.}




\maketitle
\section{Introduction}\label{sec1}

Observing a plethora of the surprising charmonium-like states named $XYZ$ states, which do not fit the quarkonia interpretation at the $B$ factories and the Tevatron, has been a new improvement in high energy physics and highlighted the need for a more comprehensive theoretical understanding of the hadron spectrum. All $XYZ$ states are above or at least in the vicinity of open-charm thresholds. So, we expect them to decay into open-charm states because of the OZI rule dominantly. But, many have only decayed to charmonium and light mesons/photons. Considering these unusual properties,
the $ Y $-states are regarded as promising candidates for unconventional hadron states. Their production mechanism, spin-parity assignments, masses, decay constants, widths, and modes have been widely discussed in review papers~\cite{Chen:2022asf,Brambilla:2019esw,Liu:2013waa,Chen:2016qju,Agaev:2020zad}. 

Over the past several decades, many charmonium-like exotic candidates have been announced, such as the scalar fully open-flavor structure $ X_{0}(2900)$~\cite{Albuquerque:2020ugi,Agaev:2022eeh}, hidden charm-strange scalar meson $X(3960)$ \cite{Agaev:2022pis,Mutuk:2022ckn}, scalar $X(5568) $ with four different quark flavors \cite{Agaev:2016urs, Sungu:2019ybf}, axial-vector hidden charm-strange state $Z_{cs}(3985) $~\cite{Maiani:2021tri,Wang:2020rcx,Sungu:2020zvk}, axial-vector hidden charm $ X(3872) $ \cite{Mutuk:2018zxs,Sundu:2016oda}, axial-vector charmed-strange state $ X(4140)$ \cite{Agaev:2017foq, Turkan:2017pil}, vector charmed-strange $X(4630)$ \cite{Agaev:2022iha}, vector hidden bottom-strange $ Y_b(10890)$ \cite{Ali:2013xba,Sungu:2018iew}, pseudo-vector meson $Z_c(3900)$ and its excited state $Z(4430)$ ~\cite{Azizi:2020yhs,Agaev:2017tzv}, etc.

Possible interpretations include meson molecules, tetraquarks, glueballs, hybrid
mesons and pentaquarks; but the status of these exotic particles is not yet fully clarified. Thus, searches of all production and decay
patterns plus the measurement of resonance parameters from other experiments, and also calculations from various theoretical studies are worthwhile in giving information that will help reveal the nature of $XYZ$ states~\cite{Chen:2022asf,Brambilla:2019esw,Liu:2013waa,Chen:2016qju,Agaev:2020zad,Wang:2021gml,Tiwari:2022azj, Liu:2021xje}.

Specifically interesting among them are the vector mesons detected in $ e^{+}e^{-} $ annihilation. Recently, a series of charmonium-like resonances with $J^{PC}=1^{--}$, so-called $ Y $ states, have been reported by countless experiments.
Still, the inner structure of the $Y$ states remains
uncertain. For instance, the state $Y(4230)$, earlier named $Y(4260)$, is the first detected vector charmonium-like state in the invariant-mass spectrum
of $\pi^{+}\pi^{-}J/\psi$ in 2005 by BaBar detector~\cite{BaBar:2005hhc}. Also, it decays into $ \pi^0\pi^0J/\psi $,
and $ K^{+}K^{-}J/\psi$ were reported in a study of $ 12.6 pb^{-1} $
data collected at $ 4.26 $ GeV by the CLEO-c experiment~\cite{CLEO:2006ike} and confirmed by Belle~\cite{Belle:2007dxy}. 

Recently, $Y(4230)$ and $Y(4500)$ have been observed by BESIII Collaboration in $ e^{+}e^{-}\rightarrow  K^{+}K^{-}J/\psi$ annihilation~\cite{BESIII:2022joj}. The second new structure is detected for the first time with a statistical significance greater than $ 8\sigma $, called $ Y(4500) $. The mass and width of the newly observed resonance are measured to be:
\begin{eqnarray}\label{MassWidthYstates}
M_{Y(4500)}&=&4484.7\pm13.3\pm24.1~\mathrm{MeV},\nonumber \\
\Gamma_{Y(4500)}&=&111.1\pm30.1\pm15.2~\mathrm{MeV}.
\end{eqnarray}
and also in the $e^{+}e^{-}\rightarrow D^{\ast0}D^{\ast-}\pi^+ + c.c$ process the following results are obtained \cite{BESIII:2023unv}:
\begin{eqnarray}\label{MassWidthYstates2023}
M_{Y(4500)}&=&4469.1\pm26.2\pm3.6~\mathrm{MeV},\nonumber \\
\Gamma_{Y(4500)}&=&246.3\pm36.7\pm9.4~\mathrm{MeV}.
\end{eqnarray}
Hadronic molecules attract special attention since many uncommon resonances are located close to the threshold of a pair of hadrons. For the $D_s\bar{D}_{s1}$ system with $J^{PC}=1^{--}$, Peng et al. \cite{Peng:2022nrj} found a bound state at $4.494$ GeV, which is consistent with the mass and quantum numbers
of the $Y(4500)$ observed recently by BESIII~\cite{BESIII:2022joj}. They have considered the $Y(4230), Y (4360), Y (4500)$ and $Y (4620)$ resonances from a molecular perspective by formulating a contact-range effective field theory for the $D\bar{D}_1$ and  $D_s\bar{D}_{s1}$ family of $S$- and $P$- wave charmed meson-antimeson systems. The $Y (4500)$ and $Y (4620)$ are estimated as hidden-strange partners of the $Y (4230)$ and $Y (4360)$, respectively. 

Additionally, in Ref.~\cite{Dong:2021juy}, charm-strange meson pairs in the
energy range relevant for the $B^+\rightarrow J/\psi\phi K^+$ is studied, and $Y (4500)$ is predicted in the threshold $D_s(1968) \bar {D}_{s1}(2536)$. This issue will become clear due to other theoretical analyses and more precise experiments. 

This paper is organized as follows: Detailed information on the QCDSR is given in Section 2. In Section 3, the results are presented and compared with other works. Finally,
we discuss our findings in Section 4.
\section{Molecule Assignment for the $Y(4500)$ via QCDSR}\label{sec:Theory}
%
QCDSR method is an entirely analytical tool and one of the most active nonperturbative models in Quantum Chromodynamics (QCD) for obtaining valuable information on hadrons \cite{Shifman:1978bx}. The main idea of the QCDSR approach is to equalize a theoretical expression of a relevant correlator with a phenomenological one. The underlying concept of this technique is \textit{duality}, which  builds a relation between hadronic (physical) and quark-gluon (QCD) degrees of freedom. The sum rules obtained allow us to compute observable properties of the hadronic state.

The starting point of the QCDSR method is the two-point correlation function, and it is determined as the vacuum expectation value of the time-ordered product of interpolation currents as stated below:
\begin{equation}\label{CorrFunc}
\Pi_{\mu \nu }(q)=i\int d^{4}x~e^{iq\cdot x}\bra 0T [j_{\mu}(x)j_{\nu}^{\dagger}(0)]\ket 0,
\end{equation}
where $T$ denotes the time-ordered product operator, $j_{\mu}$ represents the colourless local current of the $Y$ state with the quantum number $ J^{PC}=1^{--} $ and it is chosen to be \cite{Chen:2021kol}
\begin{eqnarray}\label{MoleculeCurrent}
	j_{\mu}(x)=\frac{i}{\sqrt{2}}\bigg(\bar{ c}_a(x)\gamma_{\mu}{\gamma_5}{s_a(x)}\bar   {s}_b(x){\gamma_5}{c_b(x)}
	-\bar{s}_a(x)\gamma_{\mu}{\gamma_5}{c_a(x)}\bar{c}_b(x){\gamma_5}s_b(x)\bigg),
\end{eqnarray}
where $a,b$ are colour indices. Chosen current $j_{\mu}(x)$ must contain all the information, like quark contents and quantum numbers of the examined meson. 
\subsection{Physical (Phenomenological) Side}\label{sec:Physical}
To extract the mass and decay constant QCDSR expressions, we first compute the correlation function in terms of the hadronic degrees of freedom. Inserting the complete set of hadronic states with the same quantum numbers and performing integral over $ x $ in Eq.~(\ref{CorrFunc}) we get:
\begin{eqnarray}\label{eq:denklem3}
\Pi^{Phys.}_{\mu\nu}(q)=\frac{\bra 0 j_{\mu}(0) \ket{Y(q)} \bra{Y(q)}{j}_{\nu}^{\dagger}(0) \ket 0}{m^{2}_Y-q^{2}}+subtracted~terms,
\end{eqnarray}
where $m_{Y}$ denotes the mass of $Y$
meson. The coupling of the state, Y, to the current,
$ j_{\mu} $, can be parametrized in terms of the decay constant $ f_Y $ as:
\begin{equation}\label{matrix element}
\bra 0 j_{\mu}(0)Y(q)\rangle=f_Y m_Y{\varepsilon_{\mu}},
\end{equation}
where $\varepsilon _{\mu}$ is the polarization vector of the resonance $Y$
satisfying
\begin{eqnarray}\label{polarizationvector}
	\varepsilon_{\mu}\varepsilon_{\nu}^{*}=-g_{\mu\nu}+\frac{q_{\mu
		}q_{\nu }}{m_{Y}^{2}}.
\end{eqnarray}
After employing polarization relations,
the correlation function is written in terms of Lorentz structures in the form
\begin{equation}\label{physside}
	\Pi _{\mu \nu }^{Phys.}(q)=\frac{m_{Y}^{2}f_{Y}^{2}} {%
		m_{Y}^{2}-q^{2}} \left( -g_{\mu \nu }+\frac{q_{\mu }q_{\nu }}{m_{Y}^{2}%
	}\right) +\ldots,
\end{equation}
here dots represent the contributions originating from the continuum and
higher states. Any chosen Lorentz structure's coefficient can be employed to derive the required sum rules. Here, the structure
$g_{\mu\nu}$ is selected to find the sum rules and the
typical Borel transformation in terms of
$q^2$ is performed to suppress the continuum. In the end, we get the physical side as
\begin{eqnarray}\label{Borelphys}
&&\mathcal{B}\Pi^{Phys.}=m_{Y}^{2}f_{Y}^{2}~e^{-m_{Y}^{2}/M^{2}},
\end{eqnarray}
where $M^2$ shows the auxiliary Borel parameter and choosing the scalar function for the physical side is:
\begin{equation}\label{physside2}
\Pi^{Phys.}(q^2)=\frac{m_{Y}^{2}f_{Y}^{2}} {m_{Y}^{2}-q^{2}}.
\end{equation}
%
\subsection{QCD (Theoretical) Side}\label{sec:QCDSide}
We represent hadrons by their interpolating quark currents instead of model-dependent parameters in this part. A correlation function accompanies these currents and is treated with the operator product expansion (OPE), separating the short and long-distance quark-gluon interactions. The former is handled by QCD perturbation theory, while the latter are parametrized according to universal vacuum condensates.

To evaluate the theoretical part, the current expression in Eq.\ (\ref{MoleculeCurrent})
is replaced into the correlation function given in Eq.\
(\ref{CorrFunc}). After the heavy and light quark fields are contracted, we have
the correlation function $ \Pi _{\mu \nu }^{\mathrm{QCD}}(q) $ in the molecular picture for the $Y$ state:
\begin{eqnarray}\label{eq:QCDTr}
\Pi _{\mu \nu }^{\mathrm{QCD}}(q)&=&\frac{i}{2}\int d^{4}xe^{iq\cdot x}\big\{\mathrm{Tr}[\gamma_{5}S_{c}^{bb^{\prime}}(x)\gamma_{5}S_{s}^{b^{\prime}b}(-x)]
\mathrm{Tr}[\gamma_{\mu}\gamma_{5}S_{s}^{aa^{\prime}}(x)\gamma_{5}\gamma_{\nu}S_{c}^{a^{\prime}a}(-x)]\nonumber \\
&-&\mathrm{Tr}[\gamma_{5}S_{c}^{ba^{\prime}}(x)\gamma_{5}\gamma_{\nu} S_{s}^{a^{\prime}b}(-x)]\mathrm{Tr}[\gamma_{\mu}\gamma_{5}S_{s}^{ab^{\prime}}(x)\gamma_{5}S_{c}^{b^{\prime}a}(-x)]\nonumber \\&-&\mathrm{Tr}[\gamma_5 S_{s}^{ba^{\prime}}(x)\gamma_5\gamma_{\nu}S_{c}^{a^{\prime}b}(-x)]\mathrm{Tr}[\gamma_{\mu}\gamma_{5}S_{c}^{ab^{\prime}}(x)\gamma_{5} S_{s}^{b^{\prime}a}(-x)]\nonumber \\
&+&\mathrm{Tr}[\gamma_{5}S_{s}^{bb^{\prime}}(x)\gamma_{5}S_{c}^{b^{\prime }b}(-x)]\mathrm{Tr}[ \gamma_{\mu}\gamma_{5} S_{c}^{aa^{\prime }}(x)\gamma_{5}\gamma_{\nu}S_{s}^{a^{\prime }a}(-x)]\},
\end{eqnarray}
where $S_{c,s}^{ij}(x)$ are the charm and strange quark propagators. These
quark propagators are described in connection with the quark and gluon
condensates \cite{Reinders:1984sr} and can be written down in
coordinate space using the light and heavy-quark propagators in Eqs. (\ref{eq:light-quark propagator}) and (\ref{eq:heavy-quark propagator}):
\begin{eqnarray}\label{eq:light-quark propagator}
	S_{s}^{ij}(x)&=&i\delta _{ij}\frac{\slashed x}{2\pi ^{2}x^{4}}-\delta _{ij}
	\frac{m_{s}}{4\pi ^{2}x^{2}}-\delta _{ij}\frac{\langle \overline{s}s\rangle
	}{12} +i\delta _{ij}\frac{\slashed xm_{s}\langle \overline{s}s\rangle }{48}%
	-\delta _{ij}\frac{x^{2}}{192}\langle \overline{s}g\sigma Gs\rangle  \notag \\
	&+&i\delta_{ij}\frac{x^{2}\slashed xm_{s}}{1152}\langle \overline{s}g\sigma Gs\rangle
-i\frac{gG_{ij}^{\alpha \beta }}{32\pi ^{2}x^{2}}\left[ \slashed x{\sigma
		_{\alpha \beta }+\sigma _{\alpha \beta }}\slashed x\right] -i\delta _{ij}%
	\frac{x^{2}\slashed xg^{2}\langle \overline{s}s\rangle ^{2}}{7776}  \notag \\
	&-&\delta _{ij}\frac{x^{4}\langle \overline{s}s\rangle \langle
		g^{2}G^2\rangle }{27648}+\ldots,  \label{eq:qprop}
\end{eqnarray}
\begin{eqnarray}\label{eq:heavy-quark propagator}
S_{c}^{ij}(x)&=&i\int \frac{d^{4}k}{(2\pi )^{4}}e^{-ik\cdot x} \Bigg \{ \frac{\delta _{ij}( {\slashed k}+m_{c}) }{k^{2}-m_{c}^{2}}  -\frac{gG_{ij}^{\alpha \beta }}{4}\frac{\sigma _{\alpha \beta }({\slashed k}+m_{c})+({\slashed k}+m_{c}) \sigma_{\alpha\beta }}{(k^{2}-m_{c}^{2})^{2}} \notag \\
&+&\frac{g^{2}G^{2}}{12}\delta _{ij}m_{c}\frac{k^{2}+m_{c}{\slashed k}}{(k^{2}-m_{c}^{2})^{4}}+\frac{g^{3}G^{3}}{48}\delta _{ij}\frac{( {\slashed k}+m_{c}) }{(k^{2}-m_{c}^{2})^{6}}[ {\slashed k}( k^{2}-3m_{c}^{2}) \notag \\&+&2m_{c}(2k^{2}-m_{c}^{2})] ({\slashed k}+m_{c})+\ldots\Bigg\}, 
\end{eqnarray}
where $G_{ij}^{\alpha \beta}\equiv G_{A}^{\alpha \beta}t_{ij}^{A}$
is the external gluon field strength tensors, $t_{ij}^{A}=\lambda_{ij}^{A}/2$ with
$\lambda_{ij}^{A}$ Gell-Mann matrices, $A= 1,2,...8$ shows 
color indices, $m_{s}$ and $m_{c}$ symbolizes the strange and charm quark masses respectively, $\langle \bar{s}s\rangle $
is the strange quark condensate. 
To get rid of the contributions emerging from higher states, 
the standard Borel transformation is applied to the invariant amplitude $\Pi^{QCD}(q^2)$, selecting the same structure in $\Pi^{Phys.}(q^2)$, one gets the QCD side:
\begin{equation}\label{Dispersion Integral}
\Pi^{\mathrm{QCD}}(M^2,s_{0})=\int_{s_{min}}^{s_{0}}\rho^{\mathrm{QCD}}(s)e^{-s/M^{2}}ds+\Pi(M^2),
\end{equation}
where $s_{min}= 4(m_c + m_s)^2$ and effective threshold  $s_{0}$ shows the mass of the first excited state with the same quantum number as the selection of interpolating currents for the considered particle. Based on the
analyticity, spectral density can be described as the imaginary part of the correlator by the dispersion relation as $\rho^{QCD}(s)=(1/\pi) Im[\Pi^{QCD}]$, and we evaluate up to operator dimension ten. The spectral density can be separated according to operator dimensions as
\begin{eqnarray}\label{Rhoall}
\rho^{\mathrm{QCD}}(s)&=&\rho^{\mathrm{pert.}}(s)+\sum_{N=3}^{8}\rho^{DimN}(s),
\end{eqnarray}
where $ \rho^{\mathrm{pert.}}$ denotes the spectral densities of the perturbative part and the second term implies contributions through the operator dimensions from three to eight, respectively. We do not give the explicit forms of spectral density for shortness here. Also, $\Pi(M^2) $ in Eq.~(\ref{Dispersion Integral}) expresses some non-perturbative contributions:
\begin{eqnarray}\label{PiM2}
\Pi(M^2)&=&\sum_{N=6}^{10}\Pi^{DimN}(M^2).
\end{eqnarray}
By matching the coefficients of
chosen structure, $g_{\mu \nu}$, in both physical and QCD
sides, and employing the quark hadron duality ansatz, the final decay constant sum
rule for $Y$ meson is extracted as
\begin{equation}\label{Decay constant SR}
f_{Y}^{2}=\frac{e^{m_{Y}^{2}/M^{2}}}{m_{Y}^{2}}\Pi^{QCD}(M^2,s_{0}).
\end{equation}
To obtain the final form of the mass and sum rule, the hadronic
decay constant should be removed from the sum rule expression in Eq.\ (\ref{Decay constant SR}). Leaving $\Pi^{QCD}(M^2,s_{0})$ alone in Eq.\ (\ref{Decay constant SR}) and  taking the derivative in terms of $d/d(-1/M^{2})$, the mass sum rule is obtained: 
\begin{equation}\label{massSR}
m_{Y}^{2}=\frac{\Pi^{QCD^{\prime}}(M^2,s_{0})}{\Pi^{QCD}(M^2,s_{0})},
\end{equation}
where $s_0$ is the continuum threshold and  
\begin{equation}
\Pi^{QCD^{\prime}}(M^2,s_{0})=\frac{d\Pi^{QCD}(M^2,s_{0})}{d(-1/M^{2})}.
\end{equation}
\section{Numerical Analysis}\label{sec:NumAnal}
Now, it is time to perform the numerical analysis of the sum rules
in Eqs.~(\ref{Decay constant SR}) and (\ref{massSR}). To this aim, the input parameters and condensate values are presented in Table~\ref{Table:inputparam}. 
\begin{table}[h!]
	\centering
	\caption{Input parameters}\label{Table:inputparam}
	\label{tab:Param}
	\begin{tabular}{|c|c|}
		\hline\hline
		Parameters & Values \\ \hline\hline
		$m_{c}$                                     & $1.27\pm 0.02~\mathrm{GeV}$ \cite{ParticleDataGroup:2022pth}\\
		$m_{s}$                                     & $93_{3.4}^{+8.6}~\mathrm{MeV}$ \cite{ParticleDataGroup:2022pth}\\
		$\langle \bar{s}s \rangle (1\mbox{GeV})$    & $0.8\times(-0.24\pm 0.01)^3~\mathrm{GeV}^3$ \cite{Shifman:1978bx}  \\
		$m_{0}^2 $                                  & $(0.8\pm0.1)$ $\mathrm{GeV}^2$ \cite{Dosch:1988vv}\\
		$\langle \frac{\alpha_s}{\pi} G^2 \rangle $ & $(0.012\pm0.004)$ $~\mathrm{GeV}^4 $\cite{Shifman:1978bx}\\
		$\langle g_s^3 G^3 \rangle $                & $ (0.57\pm0.29)$ $~\mathrm{GeV}^6 $\cite{Narison:2015nxh}\\
		\hline\hline	
	\end{tabular}
\end{table}
Then, the working regions of the obtained sum rules will be analysed. 
For this, we should determine the working interval of the parameters $s_0$
and $M^2$. The dominance of pole
contribution and convergence of OPE must be satisfied within the working interval of
$s_0$ and $M^2$. Note that the sum rules results must be independent of slight variations of
these parameters. 

\begin{figure}[htbp]
	\begin{minipage}[b]{0.47\linewidth}
		\centering
		\includegraphics[width=\textwidth]{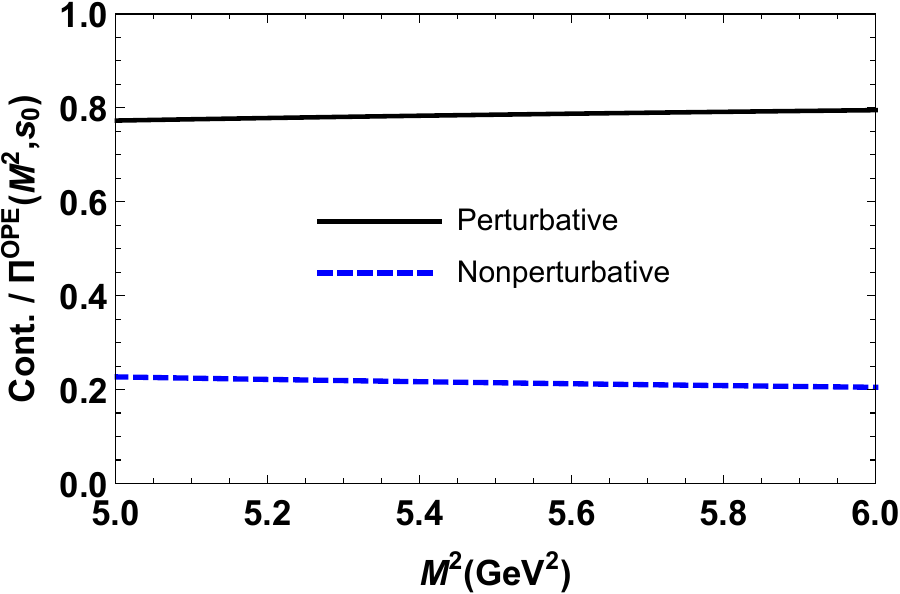}
		\caption{The relative contributions of perturbative and non-perturbative part of the OPE side for the resonance $Y(4500)$.}
		\label{fig:PertNonPertvsMsq}
	\end{minipage}
	\hspace{0.5cm}
	\begin{minipage}[b]{0.47\linewidth}
		\centering
		\includegraphics[width=\textwidth]{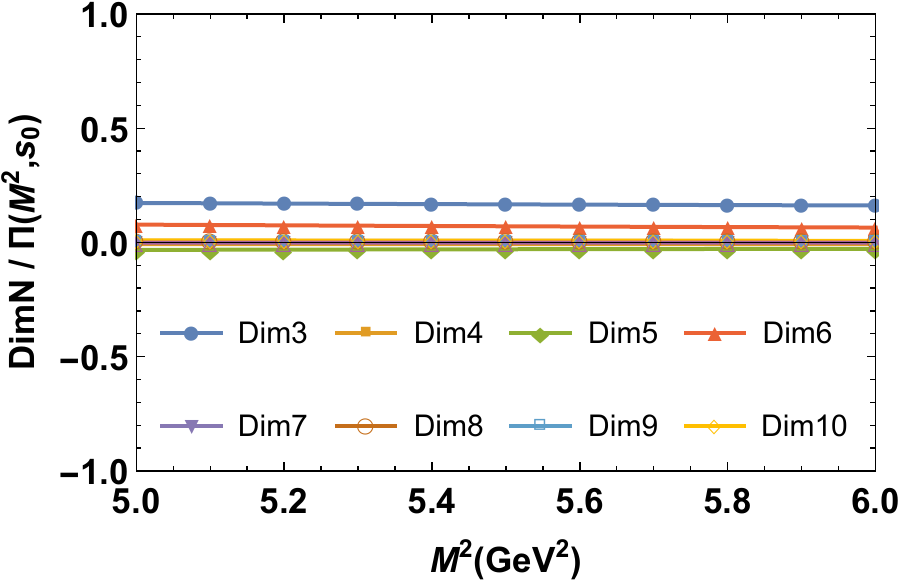}
		\caption{The relative contributions of non-perturbative part of the OPE side in terms of dimensions versus $M^2$ for the  $Y(4500)$.}
		\label{fig:Y4500NonPertDimNvsMsq}
	\end{minipage}
\end{figure}

\begin{figure}[h!]
	\begin{minipage}[b]{0.47\linewidth}
		\centering
		\includegraphics[width=\textwidth]{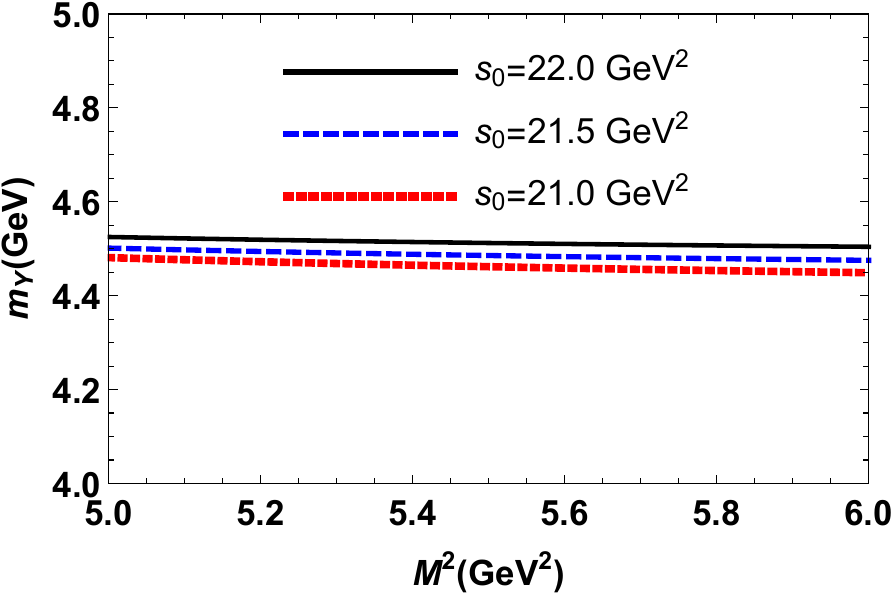}
		\caption{Mass of resonance $Y(4500)$ as a function of $M^2$. }
		\label{fig:figure1}
	\end{minipage}
	\hspace{0.7cm}
	\begin{minipage}[b]{0.47\linewidth}
		\centering
		\includegraphics[width=\textwidth]{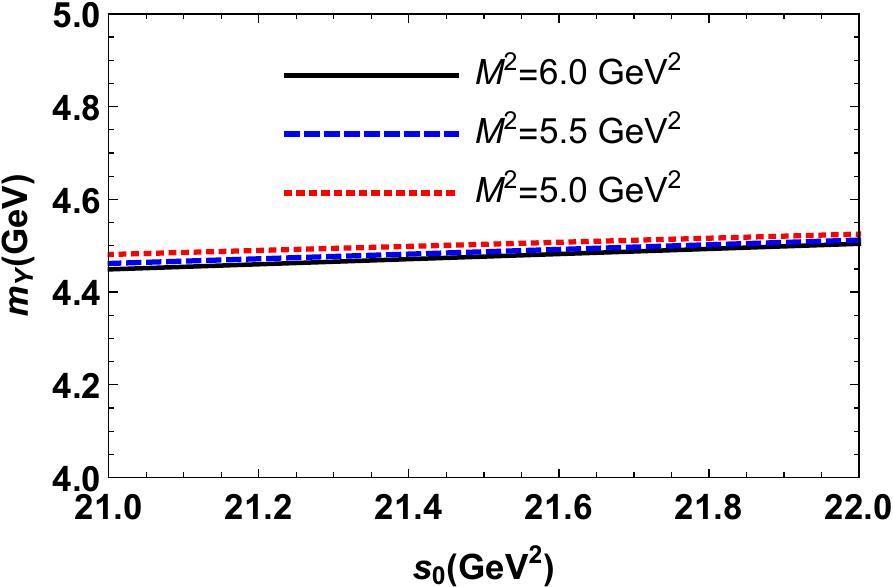}
		\caption{Mass of $Y(4500)$ state versus continuum threshold $s_0$. }
		\label{fig:figure2}
	\end{minipage}
	\hspace{0.7cm}
	\begin{minipage}[b]{0.47\linewidth}
		\centering
		\includegraphics[width=\textwidth]{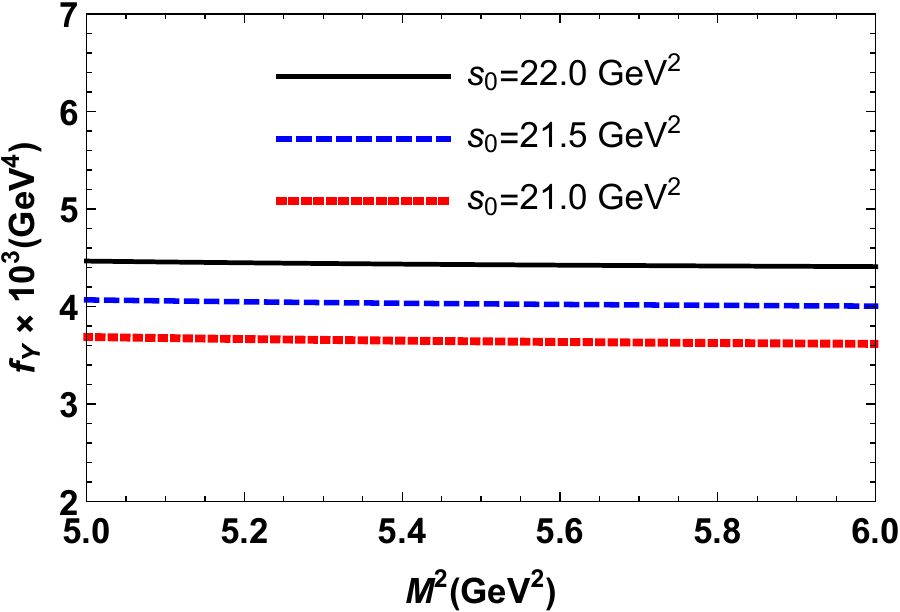}
		\caption{Decay constant of  $Y(4500)$  state as a function of Borel mass parameter $M^2$. }
		\label{fig:figure3}
	\end{minipage}
	\hspace{0.8cm}
	\begin{minipage}[b]{0.47\linewidth}
		\centering
		\includegraphics[width=\textwidth]{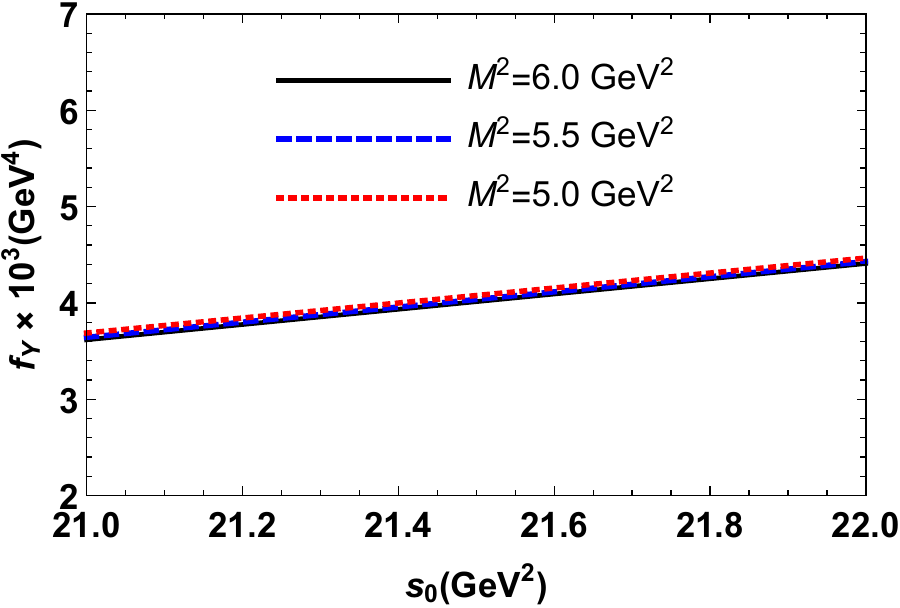}
		\caption{Decay constant of $Y(4500)$ resonance versus continuum threshold $s_0$. }
		\label{fig:figure4}
	\end{minipage}
\end{figure}

The OPE converges slowly for both tetraquark and molecular states in the QCD sum rules analysis in the literature. The lower and maximum bounds of the pole contributions in our calculation require the following condition:
\begin{equation}
\mathrm{PC}=\frac{\Pi (M_{\mathrm{max}}^{2},\ s_{0})}{\Pi (M_{\mathrm{max}}^{2},\ \infty )}>(35-73)\%.  \label{eq:Rest1}
\end{equation}
Applying the mentioned above criteria, the working regions of the
parameters $M^2$ and $s_0$ are determined as follows:
\begin{eqnarray*}\label{M2s0intervals}
	M^2 (\mathrm{GeV^2} )\in [5,6],~~~~s_0 ( \mathrm{GeV^2} )\in [21,22].
\end{eqnarray*}
Our analysis ensures that the perturbative contribution is dominant over the non-perturbative ones. In this region, the dimensions six and eight condensates become significant to compensate for the OPE convergence, as seen in Figures [\ref{fig:PertNonPertvsMsq}, \ref{fig:Y4500NonPertDimNvsMsq}] which is compatible with the selection of $s_0\simeq (m_Y+0.15$ GeV$)^2$ which is the energy of the first excited state of the related hadron. 
The analyses of mass and decay constant of the $Y$ state concerning $M^2$ and $s_0$
are plotted in Figures~[\ref{fig:figure1}-\ref{fig:figure4}] within these working regions.
Both of them are stable in terms of variations of $M^2$ and $s_0$.
\quad\\\\\\
Our results and comparison with the other theoretical studies in the literature are presented in Table \ref{tab:ResultsTable}. As seen, our mass value nicely agrees with the BESIII and effective field theory values.
\begin{table}[htbp]
	\caption{The mass and decay constant of resonance $Y(4500)$.\label{tab:ResultsTable}}
	\begin{center}
		\begin{tabular}{ |c|c|c|} 
			\hline\hline
			& $m_{Y}$                 & $f_{Y}$    \\
			& $ (\mathrm{MeV}) $    & $ (\mathrm{GeV^4}) $   \\
			\hline
			Present Work                  & $4488.35\pm11.54$          & $ (4.04\pm0.36) \times 10^{-3}  $  \\
			\hline
			Experiment                    & $4484.7\pm13.3\pm24.1$  & -  \\
			\cite{BESIII:2022joj}         &                         &    \\
			\hline
			Effective Field Theory 	      &$  4494.2_{-8.94}^{+5} $ & -	 \\
			\cite{Peng:2022nrj}           &                         &    \\
			\hline\hline	
		\end{tabular}
	\end{center}
\end{table}
\section{Conclusion}
Recent discoveries of $ XYZ $ resonances have broadened 
our perspective of the hadron spectrum and serve as excellent laboratories
to probe exotic structures. In this paper, we have focused on the charmonium-like $Y(4500 )$ state data from the $ e^{+}e^{-}\rightarrow K^{+} K^{-}J/\psi $ annihilations. We calculate the mass and decay constant of this hidden charm-strange meson with quark composition $ [c\bar{s}][\bar{c}s] $, and quantum numbers $ J^{PC}=1^{--} $ in the framework of QCDSR. We find that the $ Y(4500)$ state can be well described by the $ D_s\bar{D}_{s1} $ molecule as suggested in Refs.~\cite{Dong:2021juy, Peng:2022nrj}. More data and additional measurements are needed to provide further information about the nature of the newly observed structure and charmonium-like states.

\end{document}